\documentclass[12pt]{arXiv_class}
\usepackage{mathrsfs}

\title{\textbf{Sub-Poisson States Heralded at a Hong-Ou-Mandel Interference Peak}} 
\author{Gustavo C. Amaral, F. Calliari, T. Ferreira da Silva, G. P. Tempor\~{a}o and J. P. von der Weid}

\begin{document}
\maketitle

\begin{abstract}
A peak in coincidence events has been observed in a modified Hong-Ou-Mandel interferometer fed with weak coherent states inside the region of overlapping wave-packets. The inversion of the usual interference pattern represented by the HOM dip is equally observed when the wave packets are frequency displaced. The higher rate of coincidences indicates that photons are more likely to take on different output paths of the interferometer. This effect was harnessed so a detection on one of the arms could herald the presence of a photon in the other. The second order autocorrelation function at zero time was experimentally determined by means of the Hanburry-Brown and Twiss experiment and its value below 1 confirmed the sub-Poisson photon statistics of the heralded states.
\end{abstract}

Photon bunching is one of the most celebrated effects of two-photon interference, related to the tendency of indistinguishable photons to take the same path when there is a wave packet overlapping in a symmetric beam splitter \cite{KnightBOOK2005}. The phenomenon was first observed by Hanbury-Brown and Twiss in 1956 \cite{BrownNature1956} and later explained by Fano from a quantum mechanical perspective \cite{FanoAJP1961}. Its experimental familiarization, nevertheless, only came when Hong, Ou and Mandel demonstrated that photons generated from Spontaneous Parametric Down Conversion (SPDC) tend to bunch together when directed into a symmetric optical beam splitter \cite{HOMPRL1987}. The Hong-Ou-Mandel (HOM) \textit{dip}, which characterizes the photon bunching effect, is the drop in coincidence counts between two detectors placed at the output of the beam splitter when the indistinguishability condition is satisfied. The destructive interference involving two indistinguishable photons responsible for bunching is intimately linked to the Bose nature of photons and its atomic analog has been observed with bosonic atoms whereas the antibunching effect, intrinsically related to Fermions, was also reported with fermionic atoms \cite{JeltesNature2007}.

As we show throughout this letter, the HOM dip counterpart, the HOM \textit{peak}, can be observed in a modified version of the Hong-Ou-Mandel interferometer fed with weak coherent states. The peak is characterized by an increase in the coincidence events between detectors placed at the output of the interferometer when the wave-packets overlap. Even though this phenomenon is purely classical as it involves the interference of classical weak coherent states in a beam splitter, the higher coincidence rate indicates that, by time tuning the interferometer, it is possible to post-select states which are more likely to take on different output paths after the interaction. Following the findings of Ferreira da Silva \textit{et al.} \cite{ThiagoPRA2015}, we attempted to herald the presence of a photon in one of the output arms of the interferometer given a detection on the opposite output arm.

The quantum theory of coherence is imperative when analyzing quantum interference phenomena and permits one to determine, among many other characteristics, a light source's statistical properties \cite{MandelBOOK1995, GlauberPR1963_1}. In our case, the light source is composed of the heralded states time tuned to the HOM \textit{peak}. In the experiment of Hanburry Brown and Twiss \cite{BrownNature1956}, two detectors are placed at the output ports of a beam splitter and the coincidences are taken as a function of the delay between the temporal modes. In a heralded source context, as is our own, the detectors are gated whenever a heralding event occurs, i.e., the detections from the \textit{heralder} detector trigger both detectors (say $A$ and $B$) and the coincidence events are recorded. The value of $g^{\pars{2}}\left(0\right)$, the second-order temporal autocorrelation function at zero time, can be experimentally calculated from the rates of detections in each detector ($D_A$ and $D_B$), and the rate of coincidence events $D_{AB}$ as \cite{FaselNJP04,ThiagoPRA2015}
\begin{equation}
g^{\pars{2}}\pars{0} = \frac{D_{AB}}{D_A D_B}.
\label{eq:g2_0}
\end{equation}
The second-order temporal autocorrelation function indicates whether the photons in an optical beam are \textit{bunched} ($g^{\pars{2}}\pars{0}>g^{\pars{2}}\pars{\tau}$), \textit{antibunched} ($g^{\pars{2}}\pars{0}<g^{\pars{2}}\pars{\tau}$), or randomly distributed ($g^{\pars{2}}\pars{0}=g^{\pars{2}}\pars{\tau}$) \cite{DavidovichRMP1996}. Also, by inspecting the value of $g^{\pars{2}}\pars{0}$ calculated from Eq. \ref{eq:g2_0}, one is able to determine the ratio between multi- and single-photon pulses in the beam: when the value drops below $1$, the ratio surpasses that of a coherent field and the beam is said to be follow \textit{sub-Poisson} statistics \cite{JeltesNature2007, DavidovichRMP1996}. This result alone may also impart information regarding the antibunched nature of the photons: $g^{\pars{2}}\pars{0}<1$ implies $g^{\pars{2}}\pars{0}<g^{\pars{2}}\pars{\tau}$ and, therefore, antibunching \cite{DavidovichRMP1996}. The relation between sub-Poissonian statistics and antibunching, however, is subtle, since one can have a field with Poissonian, or even super-Poissonian, statistics which exhibits antibunching even though the converse does not apply \cite{DavidovichRMP1996}.

\begin{figure}[H]
\centering
\fbox{\includegraphics[width=0.95\linewidth]{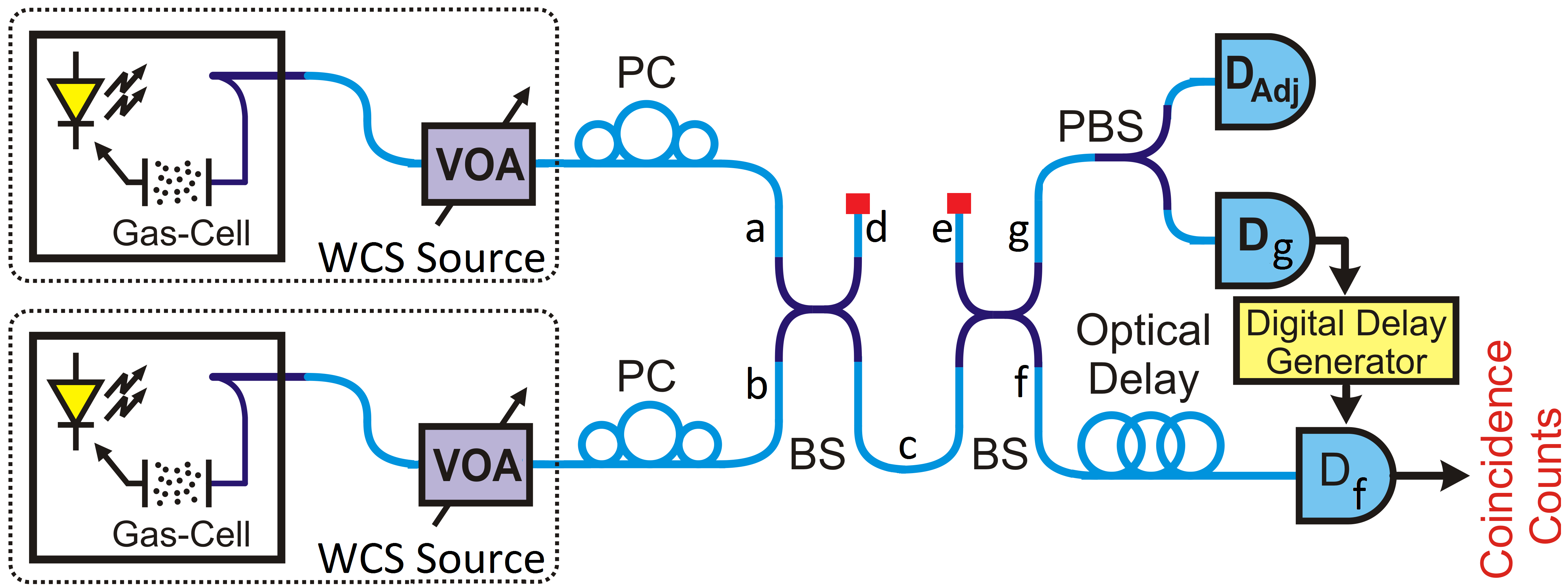}}
\caption{Modified HOM interferometer. An additional beam splitter is connected to output spatial mode $c$ of the first beam splitter and the SPADs are connected to its outputs. An electronic unit is responsible for sweeping the relative temporal delay between detections and storing the coincidence events. VOA: Variable Optical Attenuator, PBS: Polarization Beam Splitter, $D_{adj}$: polarization alignment SPAD, $D_n$: SPAD at spatial mode $n$, PC: Mechanical Polarization Controller.} 
\label{fig:ModHOM_interferometer}	
\end{figure}

Let us consider the modified Hong-Ou-Mandel interferometer fed with weak coherent states, as depicted in Fig. \ref{fig:ModHOM_interferometer}, in which a second beam splitter is connected to spatial mode \textit{c} and Single-Photon Avalanche Diodes (SPADs) operating in gated-mode are connected to spatial modes \textit{f} and \textit{g}. An electronic unit is responsible for sweeping the relative temporal delay between detections and also accumulating the coincidence events. A polarizing beam splitter (PBS) and an additional SPAD are used to guarantee that the polarization of the photons is matched: the counts in the auxiliary SPAD connected to the PBS are minimized down to the detector's noise floor. This condition is a sufficient one for the visualization of the two-photon interference effect \cite{AmaralOL2016}.

\begin{figure}[H]
\centering
\fbox{\includegraphics[width=0.95\linewidth]{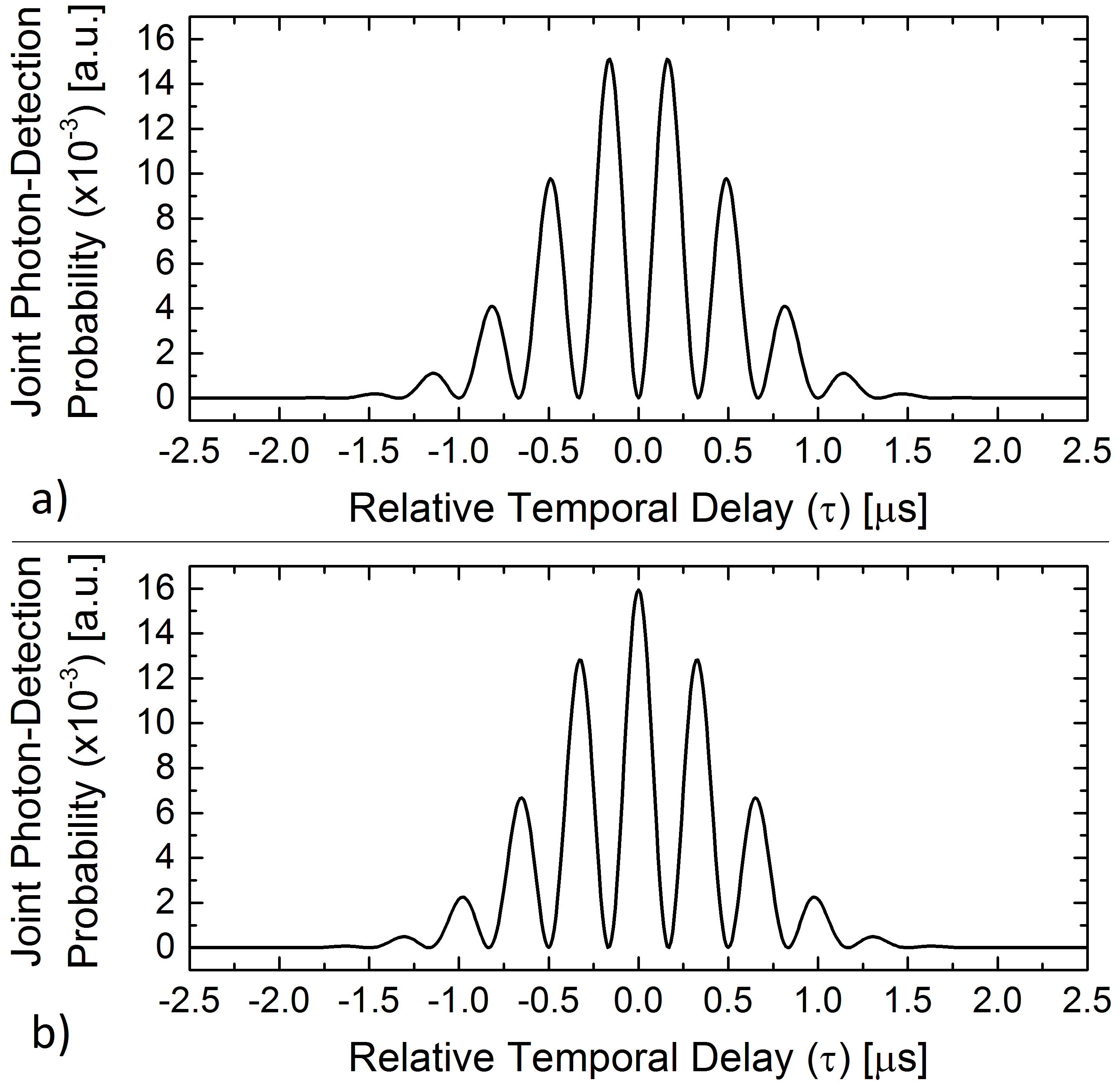}}
\caption{Numerical simulation results when gaussian wave-packets are assumed for the input states of: a) a conventional HOM interferometer; and b) the modified HOM interferometer. The probability is normalized.} 
\label{fig:SimHOMDipPeak}	
\end{figure}

The case of interest is $\ket{\psi_{in}}=\ket{1_a,1_b}=\hatd{a}_a\hatd{a}_b\ket{0}$ representing two photons entering the interferometer from input spatial modes $a$ and $b$ respectively. The spatio-temporal wave-packet formalism permits one to associate the probability of a coincidence photon detection to the wave-packet of the input states \cite{LegeroAPB2003}. Furthermore, from the expression of the joint photon-detection probability, the spectral information of the states directed to the Hong-Ou-Mandel interferometer can be extracted \cite{AmaralOL2016}. In this case, independently from the wave-packet format, the joint photon-detection probability expression recovers the vanishing character of the Hong-Ou-Mandel dip in $\tau=0$ \cite{LegeroAPB2003}. In the case of the modified Hong-Ou-Mandel interferometer, we are interested in the coincidence event probability of detectors $D_f$ and $D_g$ placed at the output of the second symmetric beam splitter. We must, therefore, attempt to write the electric field operators for modes $f$ and $g$ as a function of the input spatio-temporal wave-packets of modes $a$ and $b$. The electric field operators of modes $c$ and $d$ may be written as:
\begin{align}
\begin{aligned}
\hspace{-0.2cm}\hat{E}_c^{\pars{\!-\!}}\!\pars{\!t\!}\!=\!\tfrac{1}{\sqrt{2}}\!\pars{\!\hat{E}_a^{\pars{\!-\!}}\!\pars{\!t\!}\!+\!i\hat{E}_b^{\pars{\!-\!}}\!\pars{\!t\!}\!};
\hat{E}_d^{\pars{\!-\!}}\!\pars{\!t\!}\!=\!\tfrac{1}{\sqrt{2}}\!\pars{\!i\hat{E}_a^{\pars{\!-\!}}\!\pars{\!t\!}\!+\!\hat{E}_b^{\pars{\!-\!}}\!\pars{\!t\!}\!}.
\end{aligned}
\end{align}
The electric field operators of modes $g$ and $f$, on turn, can be written as a function of the electric field operators of modes $c$ and $e$, where we named mode $e$ the disconnected input mode of the second beam splitter of the modified Hong-Ou-Mandel interferometer:
\begin{align}
\begin{aligned}
\hat{E}_f^{\pars{-}}\pars{t}&=\tfrac{1}{\sqrt{2}}\pars{i\hat{E}_c^{\pars{-}}\pars{t}+\hat{E}_e^{\pars{-}}\pars{t}};\\
\hat{E}_g^{\pars{-}}\pars{t}&=\tfrac{1}{\sqrt{2}}\pars{\hat{E}_c^{\pars{-}}\pars{t}+i\hat{E}_e^{\pars{-}}\pars{t}}.
\end{aligned}
\end{align}
Combining these equations, we find that
\begin{align}
\begin{aligned}
\hat{E}_f^{\pars{-}}\pars{t}&=\tfrac{i}{2}\pars{\hat{E}_a^{\pars{-}}\pars{t}+i\hat{E}_b^{\pars{-}}\pars{t}}+\tfrac{1}{\sqrt{2}}\hat{E}_e^{\pars{-}}\pars{t};\\
\hat{E}_g^{\pars{-}}\pars{t}&=\tfrac{1}{2}\pars{\hat{E}_a^{\pars{-}}\pars{t}+i\hat{E}_b^{\pars{-}}\pars{t}}+\tfrac{i}{\sqrt{2}}\hat{E}_e^{\pars{-}}\pars{t}.
\end{aligned}
\label{eq:ElectricOpModHOM}
\end{align}
Once again, we are interested in the joint photon-detection probability at modes $f$ and $g$ given that two single-photons are directed each to a different input port of the modified Hong-Ou-Mandel interferometer. The expression is quite similar to the one for the usual Hong-Ou-Mandel interferometer \cite{AmaralOL2016, LegeroAPB2003}, but with a slight change of notation due to the fact that the mode names are different:
\begin{align}
\begin{aligned}
&P_{joint}\pars{t,\tau}=g_{f,g}\pars{t,t\!+\!\tau}=\\
&\bra{0}\hat{a}_a\hat{a}_b\hat{E}_f^{\pars{-}}\!\pars{t}\!\hat{E}_g^{\pars{-}}\!\pars{t\!+\!\tau}\!\hat{E}_f^{\pars{+}}\!\pars{t\!+\!\tau}\!\hat{E}_g^{\pars{+}}\!\pars{t}\!\hatd{a}_b\hatd{a}_a\ket{0}.
\end{aligned}
\end{align}
When we substitute the expression for the electric field operators on modes $f$ and $g$ as a function of the input wave-packets, $\hat{E}_{a,b}^{\pars{-}}\!\pars{t}\!=\!\varepsilon_{a,b}\!\pars{t}\hatd{a}_{a,b}$, the result yields
\begin{align}
&P_{joint}\!\pars{\!t,\tau\!}\!=\!\tfrac{1}{16}|\varepsilon_a\!\pars{\!t\!+\!\tau\!}\varepsilon_b\!\pars{\!t\!}\!+\!\varepsilon_b\!\pars{\!t\!+\!\tau\!}\!\varepsilon_a\!\pars{\!t\!}|^2.
\end{align}
One of the most interesting features of this result is that the \textit{joint photon-detection probability} reaches its maximum value at $\tau=0$, characterizing the Hong-Ou-Mandel peak. In Fig.\ref{fig:SimHOMDipPeak}-a and Fig.\ref{fig:SimHOMDipPeak}-b, we show the numerical simulation results when gaussian wave-packets are considered for the input frequency-displaced states entering a Hong-Ou-Mandel interferometer (for which the \textit{joint photon-detection probability} is given by $P_{joint}\pars{t,\tau}=\tfrac{1}{4}|\varepsilon_a\pars{t\!+\!\tau}\varepsilon_b\pars{t}\!-\!\varepsilon_b\pars{t\!+\!\tau}\varepsilon_a\pars{t}|^2$) and for the modified Hong-Ou-Mandel interferometer. The frequency displacement is set to $20$MHz and, as expected, the probability of a coincidence event is zero at $\tau=0$ and raises to a maximum at $\tau=pi/\Delta=0.16\mu$s for Fig. \ref{fig:SimHOMDipPeak}-a and approaches its peak value as $\tau$ goes to zero and goes to zero when $\tau=pi/\Delta$ for \ref{fig:SimHOMDipPeak}-b.

\begin{figure}[H]
\centering
\fbox{\includegraphics[width=0.95\linewidth]{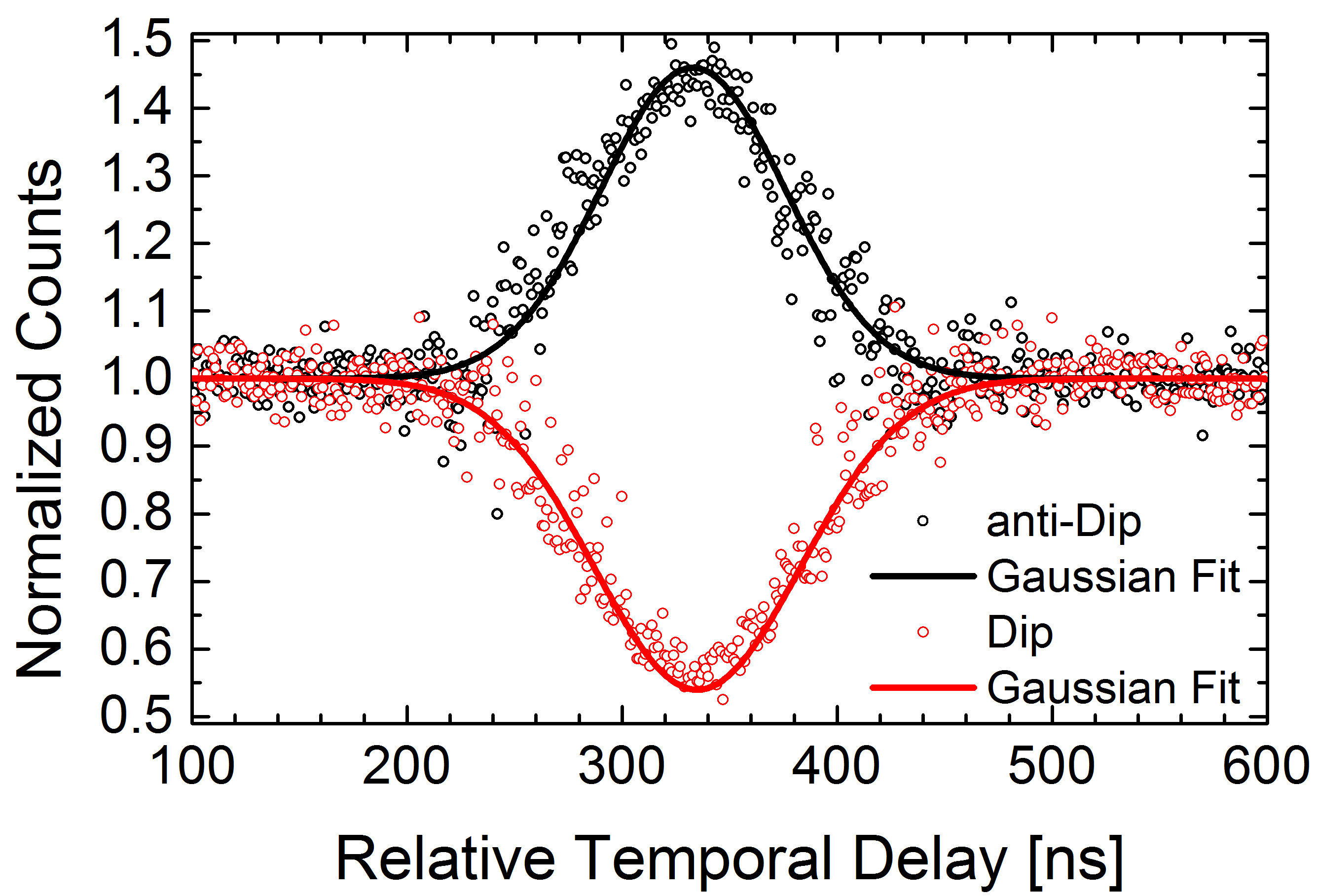}}
\caption{HOM dip (red) and  HOM peak (black) interferogram. The HOM dip is acquired by placing detectors at spatial modes $c$ and $d$, whereas the HOM peak is acquired by placing the detectors at spatial modes $f$ and $g$.} 
\label{fig:Peak_and_Dip}	
\end{figure}

To perform the experimental verification of the above theoretical predictions, the optical output of a frequency-locked CW semiconductor laser centered at $1551.7$ nm was directed to one of the input ports of the first symmetric beam splitter through a polarization controller and a variable optical attenuator. The remaining input port of the beam splitter was connected to another frequency-locked CW semiconductor laser centered at the same wavelength as the first. Frequency stabilization is possible by locking the laser wavelength to the absorption line of a high-Q factor gas-cell through a PID control loop. Polarization and intensity control was also enforced onto the second laser to guarantee the indistinguishable and few-photon conditions necessary for observation of the two-photon interference phenomenon with weak-coherent states \cite{AmaralOL2016}. The result of the experiment is presented in Fig. \ref{fig:Peak_and_Dip}, where the red curve corresponds to the normalized HOM dip acquired by placing detectors at spatial modes $c$ and $d$, and the black curve corresponds to the HOM peak when the detectors are placed at spatial modes $f$ and $g$ (refer to Fig. \ref{fig:ModHOM_interferometer}).  Both detectors are set to $15\%$ efficiency and $4$-ns wide gates which corresponds to the best compromise between dark counts (on the order of $10^{-5}$ per detection gate) and efficiency.

\begin{figure}[H]
\centering
\fbox{\includegraphics[width=0.95\linewidth]{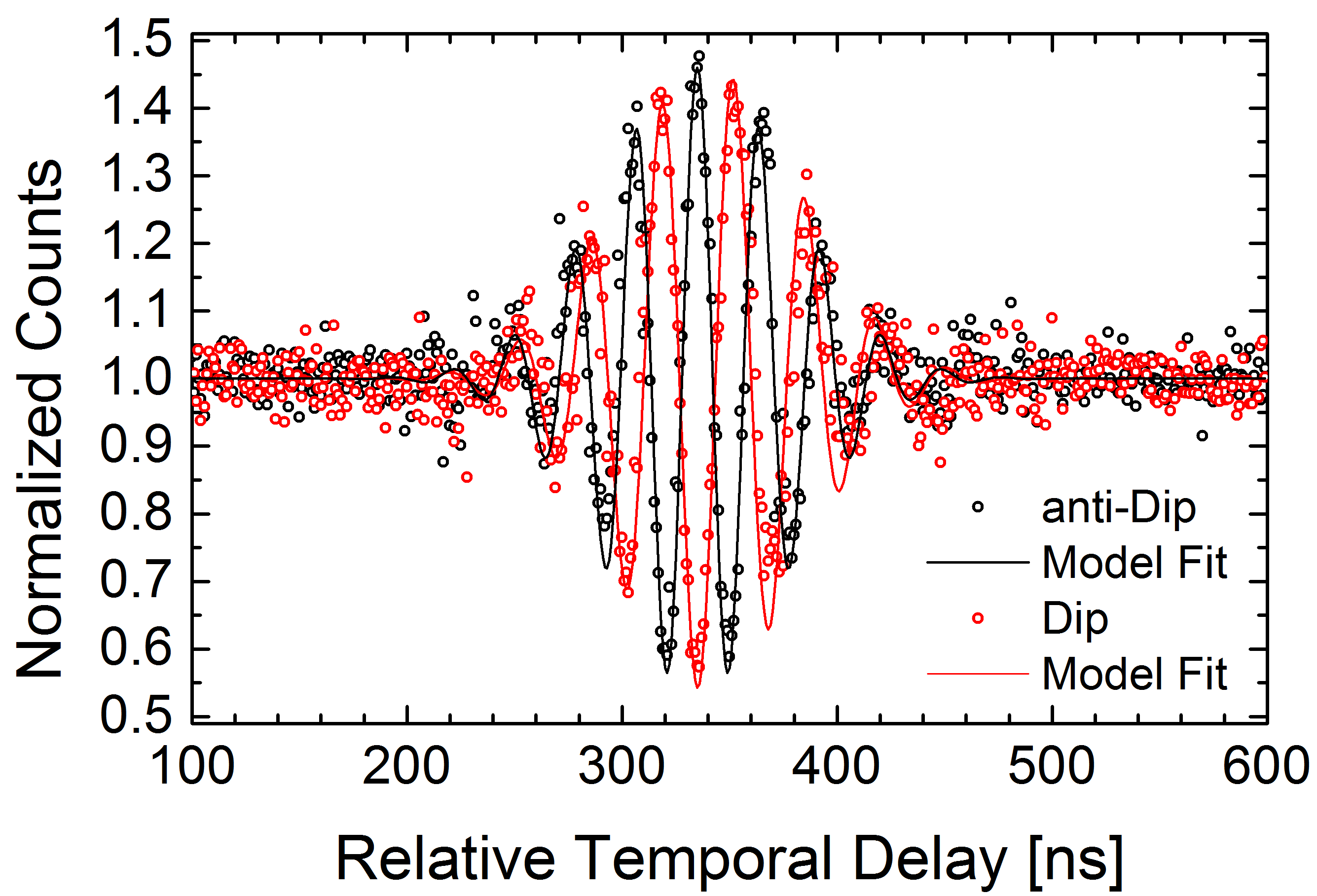}}
\caption{Beat pattern modulating the HOM dip (red) and the HOM peak (black) interferogram curves in the case of frequency-displaced wave packets.}
\label{fig:Pek_and_Dip_Beat}
\end{figure}

In order to guarantee that the relative temporal delay for both experiments was identical as observed in Fig. \ref{fig:Peak_and_Dip}, the same optical fibre loop was employed to impose an optical delay between the output of the second beam splitter and the input of the SPAD, which corresponds to $\sim 343$ ns. We see that the intrinsic multi-photon limitation associated to employing weak-coherent states in two-photon interference experiments \cite{OuBOOK2006} also affects the HOM peak, the visibility does not reach $100\%$ and is limited by $50\%$ \cite{RarityJOB2005}. In Fig. \ref{fig:Pek_and_Dip_Beat}, the interferogram for both the HOM dip and the HOM peak are presented when the frequency displacement condition is enforced. We used the model developed in \cite{ThiagoJOSAB2015} for data-fitting the interferogram with good correspondence. We also assumed gaussian wave packets and assured that the average number of photons per gate $\mu$ was kept smaller than the $0.2$ limitation of the theoretical model \cite{ThiagoJOSAB2015}.

The experimental results clearly advocate that, from a heralded source point of view, one is able to time tune the heralding event to match the Hong-Ou-Mandel peak or any other region of the interferogram. Conversely, one could also herald a state at the Hong-Ou-Mandel dip. This is important in order to compare the results of $g^{\pars{2}}\pars{0}$ for each case at the light of the quantum coherence theory. To conduct the HBT experiment and, thus, experimentally determine the value of $g^{\pars{2}}\pars{0}$, we consider the detector $D_g$ as a heralder for the presence of a photon in spatial mode $f$: we are interested in determining the multi- to single-photon ratio in this beam conditioned to a detection in $D_g$.

\begin{figure}[H]
\centering
\fbox{\includegraphics[width=0.95\linewidth]{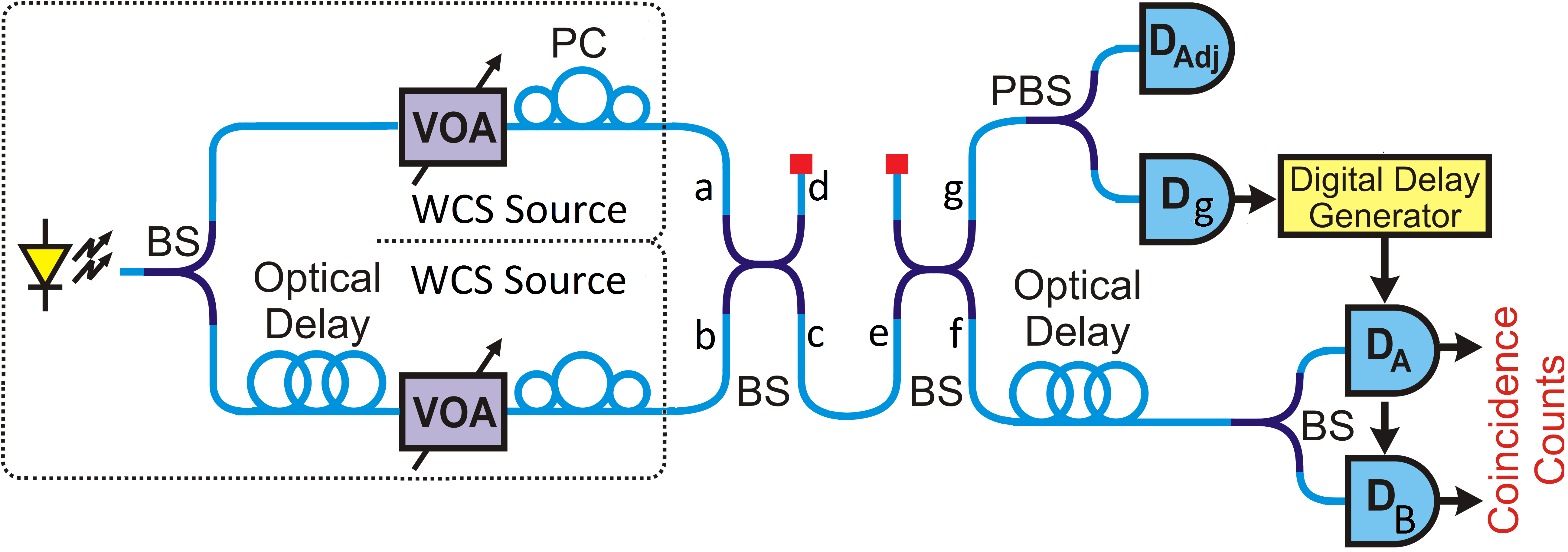}}
\caption{Simplified experimental apparatus employing a single laser source in a self-homodyne configuration. The weak coherent states are uncorrelated. The connections necessary for the HBT experiment are depicted: the connection of an additional beam splitter at spatial mode $f$; and two SPADs ($D_A$ and $D_B$) at its outputs.} 
\label{fig:SingleLaserModHOM_interferometer}	
\end{figure}

Two SPADs ($A$ and $B$) are placed at both spatial output modes of another symmetric beam splitter. The input ports of this beam splitter are arranged so that one of them is connected to spatial mode $f$ and the other is left unconnected. The experimental setup is clarified in Fig. \ref{fig:SingleLaserModHOM_interferometer}. To eliminate experimental complications due to the frequency-locking loop, we introduce a self-homodyne setup with a single laser source \cite{ThiagoPRA2015}: both frequency-locked lasers are substituted by a single semiconductor CW laser; the output is divided by an optical beam splitter and the coherent states in each arm are uncorrelated by an optical delay line approximately $8$-kilometers long to account for the temporal coherence of the source; the independent states are then directed to the modified HOM interferometer which, after the modifications, acts as a heralded source connected to a HBT experiment. The complete setup is depicted in Fig. \ref{fig:SingleLaserModHOM_interferometer}. This new setup has the advantages of employing a single laser source and the absence of frequency-locking control which greatly simplifies the experimental apparatus. Other than the fact that the new laser source employed is spectrally narrower when compared to the former frequency-locked lasers, no fundamental changes are made to the experiment. 

\begin{figure}[H]
\centering
\fbox{\includegraphics[width=0.95\linewidth]{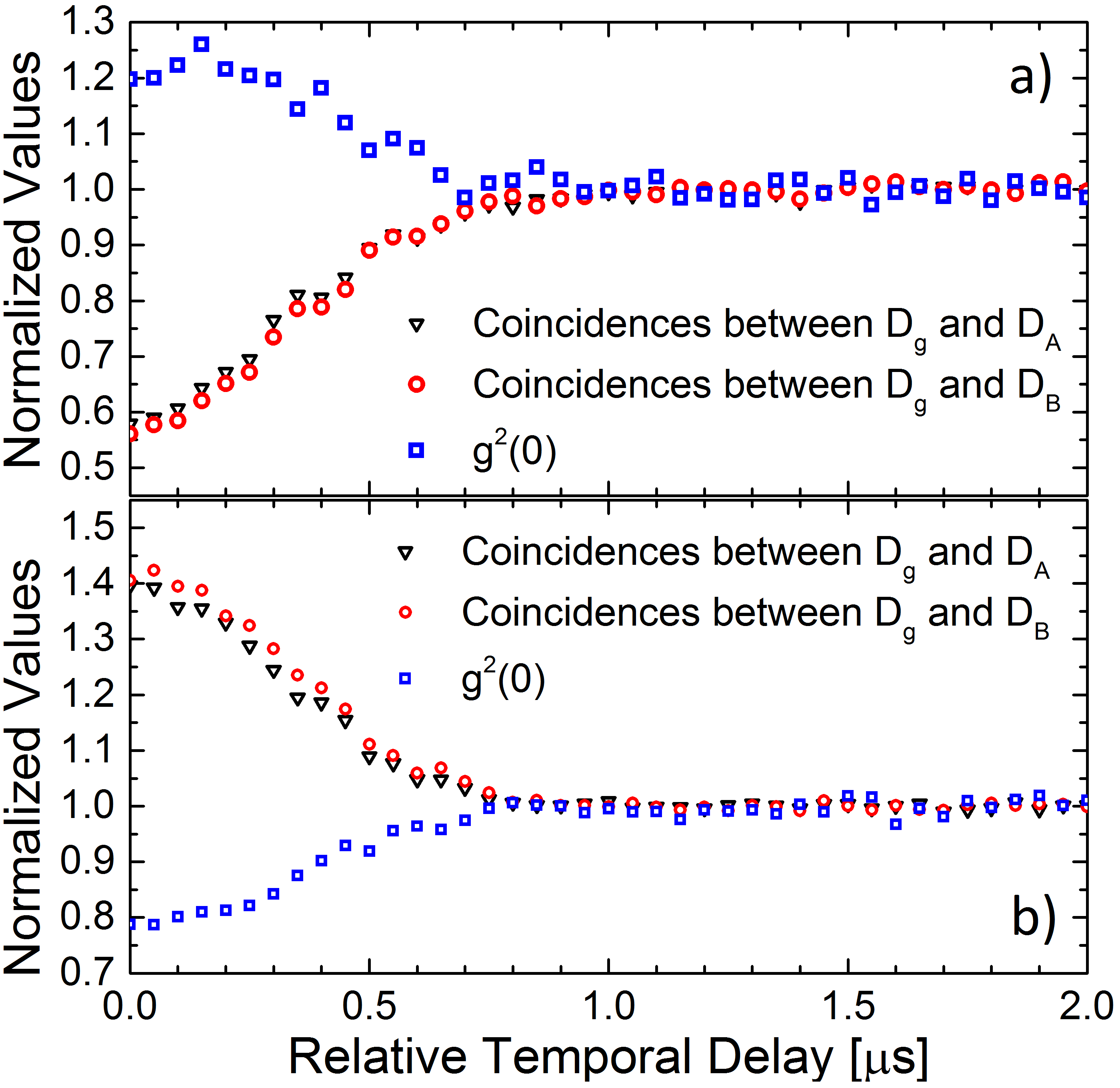}}
\caption{Experimentally determined values of $g^{\pars{2}}\pars{0}$ (blue squares) for different time-tuned heralding temporal delays in the HBT. The values are calculated for both the HOM dip (a) and the HOM peak (b), as indicated by the coincidences between $D_g$ and $D_A$ (black triangles) and between $D_g$ and $D_B$ (red circles).} 
\label{fig:g2(0)_results}	
\end{figure}

In Fig. \ref{fig:g2(0)_results}, the results of the HBT experiment are depicted. The value of $g^{\pars{2}}\pars{0}$ was calculated for different time-tuned heralding instants covering the whole region of overlapping wave-packets and also outside this region. The experiment was performed for the HOM dip (Fig. \ref{fig:g2(0)_results}-a) and for the HOM peak(Fig. \ref{fig:g2(0)_results}-a). For clarity, we plot, for each value of the experimentally determined $g^{\pars{2}}\pars{0}$, the respective accumulated counts on both $D_A$ and $D_B$ conditioned to a delayed detection in $g$. Note that these counts represent the HOM interferogram between $D_g$ and $D_A$, and between $D_g$ and $D_B$, respectively. Upon close inspection of the experimental apparatus depicted in Fig.\ref{fig:SingleLaserModHOM_interferometer} for the HBT experiment and in Fig.\ref{fig:ModHOM_interferometer}, one can identify that, apart from the intrinsic optical loss induced by the symmetric beam splitter, the setup is also capable of tracing the HOM interferogram. 

The results clearly indicate a difference in the nature of the states heralded at three distinct regions: the HOM dip; the HOM peak; and the region of distinguishable state in which the wave-packets are not overlapped. As expected, when the states are distinguishable, no variation of $g^{\pars{2}}\pars{0}$ is observed and it is tied to the value of 1 corresponding to states with Poisson photon statistics such as the coherent states. An incoherent mix of coherent states at the input of a beam splitter produce an incoherent mix of coherent states at the output so $g^{\pars{2}}\pars{0}=1$ \cite{KnightBOOK2005}. When the heralding instant matches the HOM dip, on the other hand, we see that the higher probability of states to take on the same path cause the photon statistic of the heralded states to follow a super-Poisson distribution, a result confirmed by $g^{\pars{2}}\pars{0}$ greater than 1. Conversely, in the HOM peak, the states are more likely to take on different output paths so a herald event is likely to announce a light pulse containing at least one photon and, with the condition of weak coherent states satisfied, unlikely to announce a multi-photon pulse. The photon statistics of the heralded states in this case follow a sub-Poisson distribution with $g^{\pars{2}}\left(0\right)$ less than 1.

Squeezed states of light which follow a sub-Poisson distribution are specially interesting since they approximate the single-photon condition. The fact that such states can be produced with a simple linear optic setup and a conditioned heralding detection is exciting since it eliminates the need of complicated setups involving non-linear effects with high-energy pump sources. From a quantum cryptography point of view, the results of \cite{ThiagoPRA2015} show that linear-optic heralded sub-Poisson states have similar performance when compared states heralded through the process of Spontaneous Parametric Down Conversion. Our claim that the heralding event after the two-photon interference of classical states at an optical beam splitter can indeed yield states with non-classical photon distribution, such as sub-Poisson, is grounded by the results of $g^{\pars{2}}\pars{0}=0.78$ for the HOM peak, $g^{\pars{2}}\pars{0}=1.25$ for the HOM dip, and $g^{\pars{2}}\pars{0}=1$ when no quantum interference takes place.

\begin{flushleft}
\textbf{Funding.} Work partially supported by CNPq and FAPERJ. The authors would like to thank A. H. Cordes and M. G. Albarracin for help with the frequency-locking control loop.
\end{flushleft}

\bibliographystyle{IEEEtran}
\bibliography{References}

\end{document}